\documentclass[
    ,final            
  ]
  {aipproc}

\layoutstyle{6x9}
\usepackage{wrapfig} 

\newcommand{\be}{\begin{equation}}
\newcommand{\ee}{\end{equation}}
\newcommand{\ba}{\begin{eqnarray}}
\newcommand{\ea}{\end{eqnarray}}
\newcommand{\la}{\langle}
\newcommand{\ra}{\rangle}
\newcommand{\di}{ {\rm d} }
\newcommand{\aH}{ {a} }
\def\Journal#1#2#3#4{{\it #1} {\bf #2} #3 #4} 

\begin{document}

\title{Collins Effect in SIDIS and in ${\bf e^+e^-}$ Annihilation}
\classification{13.88.+e, 13.60.Br, 13.85.Ni, 13.85.Qk}
\keywords      {QCD; partons; polarization; asymmetry; chiral model}

\author{A.~V.~Efremov}{
  address={Joint Institute for Nuclear Research, Dubna, 141980 Russia.
  E-mail: efremov@theor.jinr.ru}
}
\author{K.~Goeke }{
address={Institut f\"ur Theoretische Physik II, Ruhr-Universit\"at
Bochum, Germany}
}
\author{P.~Schweitzer}{
  address={Institut f\"ur Theoretische Physik II, Ruhr-Universit\"at
Bochum, Germany}
}
 
\begin{abstract} 
We review the present understanding of the nucleon transversity 
distribution and Collins fragmentation function, based on 
Ref.\cite{Efremov:2006qm}, and discuss how Drell-Yan experiments 
will improve it. 
\end{abstract}
\maketitle 

\paragraph{1. Introduction}
The chirally odd transversity distribution function $h_1^a(x)$ 
cannot be extracted from data on semi-inclusive deep inelastic 
scattering (SIDIS) alone. It enters the expression for the 
Collins single spin asymmetry (SSA) in SIDIS together with the 
chirally odd and equally unknown Collins fragmentation function 
\cite{Collins:1992kk} (FF) $H_1^a(z)$  \footnote{
    \label{Footnote-1}
    We assume a factorized Gaussian dependence on parton and 
    hadron transverse momenta \cite{Mulders:1995dh} with
    $B_{\rm G}(z)=(1+z^2\;\la{\bf p}_{h_1}^2\ra/\la{\bf K}^2_{H_1}\ra)^{-1/2}$ 
    and define $H_1^\aH(z) \equiv H_1^{\perp (1/2) a}(z)= $
    $\int\di^2{\bf K}_T\frac{|{\bf K}_T|}{2zm_\pi} H_1^{\perp a}(z,{\bf K}_T)$
    for brevity.
    The Gaussian widths are assumed flavor and $x$- or $z$-independent.
    We neglect throughout soft factors \cite{Ji:2004wu}.}
\be\label{Eq:AUT-Collins-1}
    A_{UT}^{\sin(\phi+\phi_S)}
    \!= 2\frac{\sum_a e_a^2 x h_1^a(x)B_{\rm G}
    H_1^{a}(z)}{\sum_a e_a^2\,x f_1^a(x)\,D_1^{a}(z)} \;.
\ee

However, $H_1^{a}(z)$ is accessible in $e^+e^-\to \bar q q\to 
2{\rm jets}$ where the quark transverse spin correlation induces 
a specific azimuthal correlation of two hadrons in opposite jets 
\cite{Boer:1997mf}
\be\label{Eq:A1-in-e+e}
    \di\sigma=\di\sigma_{\rm unp}\underbrace{\Biggl[
    1+\cos(2\phi_1)\frac{\sin^2\theta}{1+\cos^2\theta}
    \;C_{\rm G}\times
    \frac{\sum_a e_a^2 H_1^{a}H_1^{\bar a}}
    {\sum_a e_a^2 D_1^a D_1^{\bar a}}\Biggr]}_{\equiv A_1}
\ee
where $\phi_1$ is azimuthal angle of hadron 1 around z-axis along 
hadron 2, and $\theta$ is electron polar angle. Also here we 
assume the Gauss model and $C_{\rm 
G}(z_1,z_2)=\frac{16}{\pi}{z_1z_2}/{(z_1^2+z_2^2)}$.

First experimental indications for the Collins effect were 
obtained from studies of preliminary SMC data on SIDIS 
\cite{Bravar:1999rq} and DELPHI data on charged hadron production 
in $e^+e^-$ annihilations at the $Z^0$-pole \cite{Efremov:1998vd}. 
More recently HERMES reported data on the Collins (SSA) in SIDIS 
from proton target \cite{Airapetian:2004tw,Diefenthaler:2005gx} 
giving the first unambiguous evidence that $H_1^a$ and $h_1^a(x)$ 
are non-zero, while in the COMPASS experiment 
\cite{Alexakhin:2005iw} the Collins effect from a deuteron target 
was found compatible with zero within error bars. Finally, last 
year the BELLE collaboration presented data on sizeable azimuthal 
correlation in $e^+e^-$ annihilations at a center of mass energy 
of $60\,{\rm MeV}$ below the $\Upsilon$-resonance 
\cite{Abe:2005zx,Ogawa:2006bm}.

The question which arises is: {\sl Are all these data
from different SIDIS and $e^+e^-$ experiments compatible, i.e.\
due to the same effect, namely the Collins effect?}

In order to answer this question we extract $H_1^a$ from HERMES 
\cite{Diefenthaler:2005gx} and BELLE 
\cite{Abe:2005zx,Ogawa:2006bm} data, and compare the obtained 
ratios $H_1^a/D_1^a$ to each other and to other experiments. Such 
``analyzing powers'' might be expected to be weakly 
scale-dependent, as the experience with other spin observables 
\cite{Ratcliffe:1982yj,Kotikov:1997df} indicates.

\paragraph{2. Collins effect in SIDIS}
In order to extract information on Collins FF from SIDIS 
a model for the unknown $h_1^a(x)$ is needed. We use 
predictions from chiral quark-soliton model \cite{Schweitzer:2001sr}
which provides a good description of unpolarized and helicity 
distribution \cite{Diakonov:1996sr}.
On the basis of Eq.~(\ref{Eq:AUT-Collins-1}), the assumptions
in Footnote~\ref{Footnote-1}, and the parameterizations 
\cite{Gluck:1998xa,Kretzer:2001pz} for $f_1^a(x)$ and $D_1^a(z)$
at $\la Q^2\ra=2.5\,{\rm GeV}^2$, 
we obtain from the HERMES data \cite{Diefenthaler:2005gx}:
\be\label{Eq:B-Gauss-H1perp12-fav}
    \la 2B_{\rm G}H_1^{\rm fav}\ra =  (3.5\pm 0.8)\;,\quad
    \la 2B_{\rm G}H_1^{\rm unf}\ra = -(3.8\pm 0.7)\;.
\ee
Here ``${\rm fav}$'' (``${\rm unf}$'') means favored $u\to\pi^+,\,d\to\pi^-$, etc.\ 
(unfavored $u\to\pi^-$, etc.) fragmentation, and $\la\dots\ra$ denotes average over
$z$ within the HERMES cuts $0.2\le z \le 0.7$. 

\begin{figure}
\includegraphics[width=1.45in]{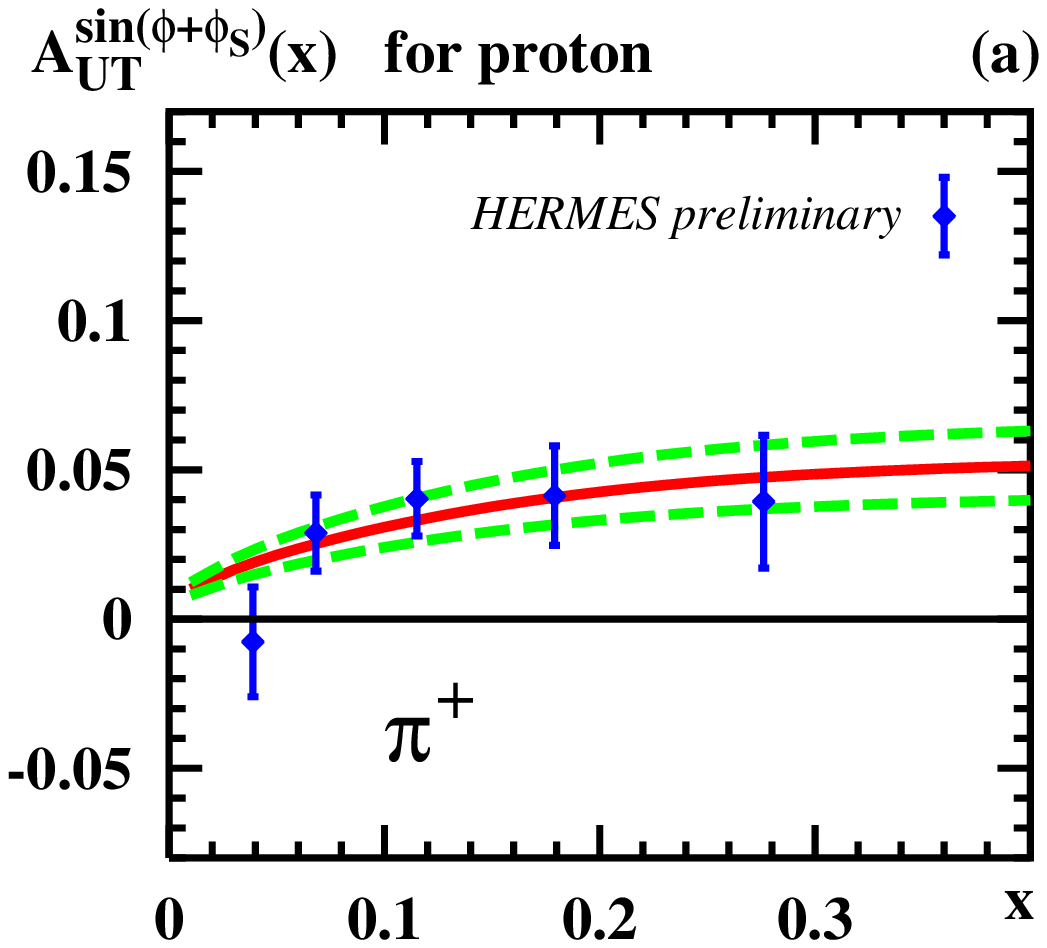}
\includegraphics[width=1.45in]{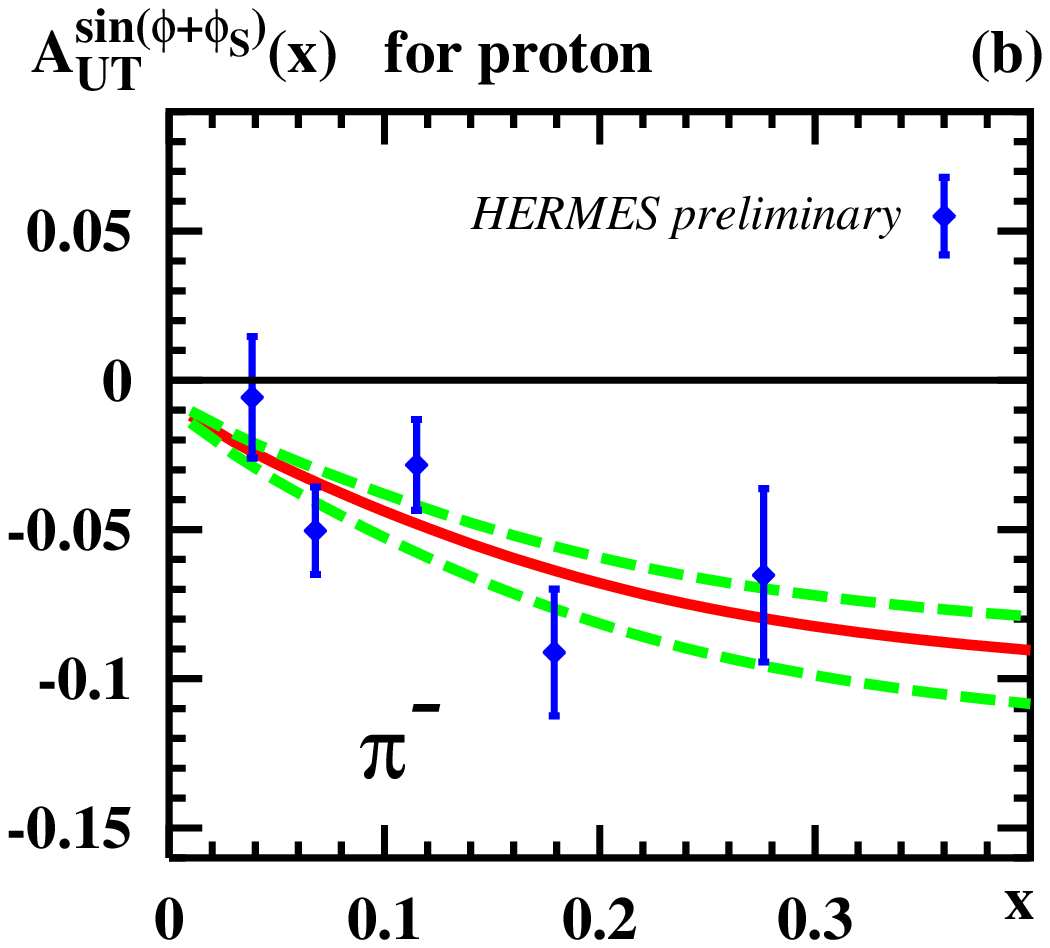}
\hspace{-5mm}
\includegraphics[width=1.59in]
{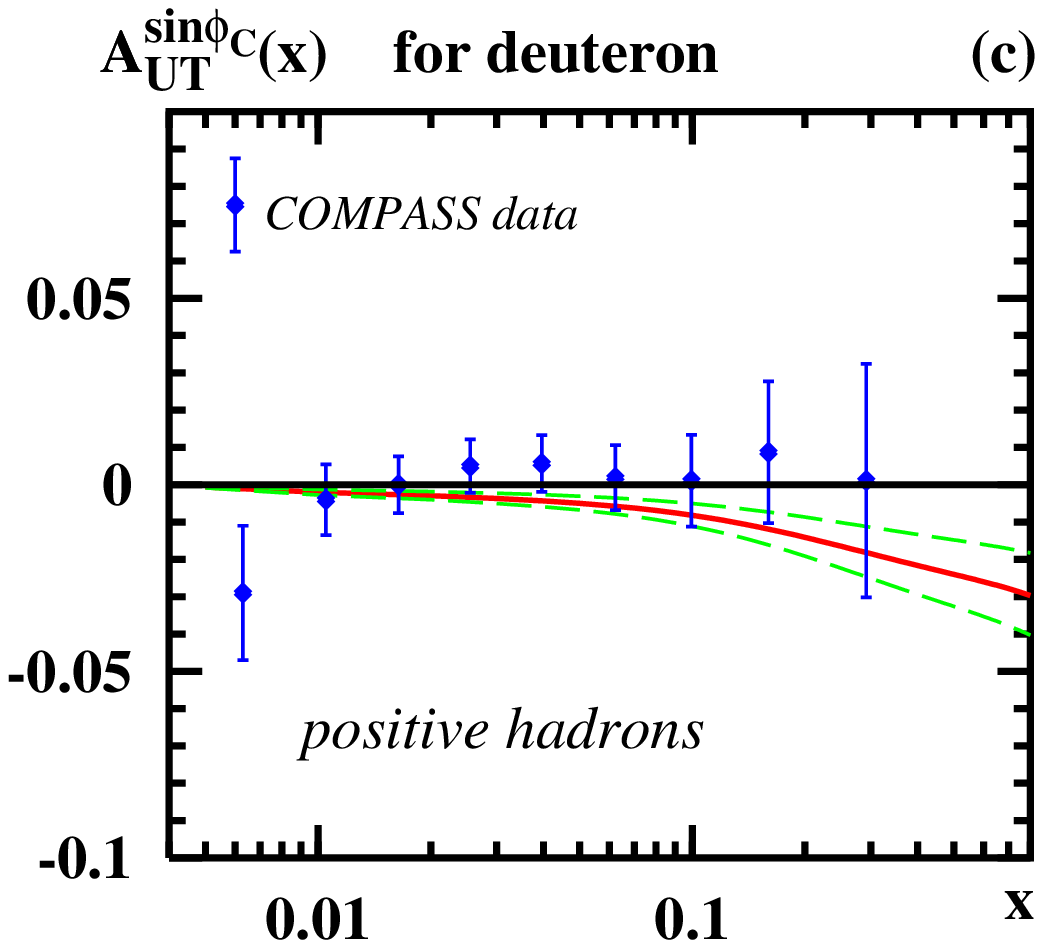}
\hspace{-5mm}
\includegraphics[width=1.59in]
{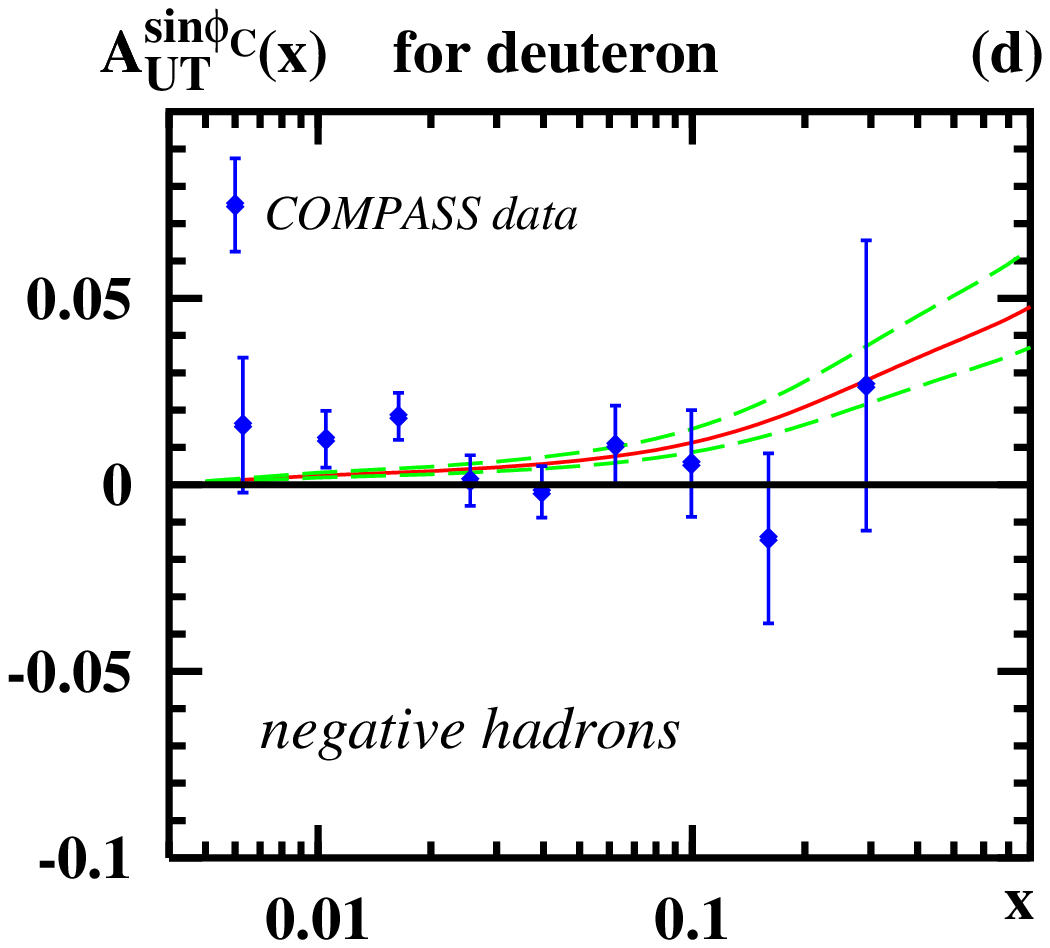}
\caption{\label{Fig3:AUT-x}
Collins SSA $A_{UT}^{\sin(\phi+\phi_S)}$ as function of $x$ vs.\   
HERMES \cite{Diefenthaler:2005gx} and new COMPASS 
\cite{Alexakhin:2005iw} data.} 
\end{figure}

Thus, the favored and unfavored Collins FFs appear to be of similar magnitude 
and opposite sign. The string fragmentation picture \cite{Artru:1995bh} 
and Sch\"afer-Teryaev sum rule \cite{Schafer:1999kn} provide a 
qualitative understanding of this behavior.
The important role of unfavored FF becomes
more evident by considering the analyzing powers 
\be
\label{Eq:Apower-HERMES}
\frac{\la 2 B_{\rm G}H_1^{\rm fav}\ra}{\la D_1^{\rm fav}\ra}
\biggr|_{\rm HERMES}\hspace{-11mm}=(7.2\pm 1.7)\% \;, \;\;\;\; \\
\frac{\la 2 B_{\rm G}H_1^{\rm unf}\ra}{\la D_1^{\rm unf}\ra}
\biggr|_{\rm HERMES}\hspace{-11mm} = -(14.2\pm 2.7)\%\;.
\ee
Fit (\ref{Eq:B-Gauss-H1perp12-fav}) describes satisfactorily 
the HERMES proton target data \cite{Diefenthaler:2005gx} on the 
Collins SSA (see Figs.~\ref{Fig3:AUT-x}a, b) and is in agreement 
with COMPASS deuteron data \cite{Alexakhin:2005iw} 
(Figs.~\ref{Fig3:AUT-x}c, d).

%

%
\begin{wrapfigure}[13]{HR}{2.1in} 
\label{Fig5:BELLE-best-fit}
\vspace{-5mm}
\begin{flushright}
\includegraphics[width=1.8in]{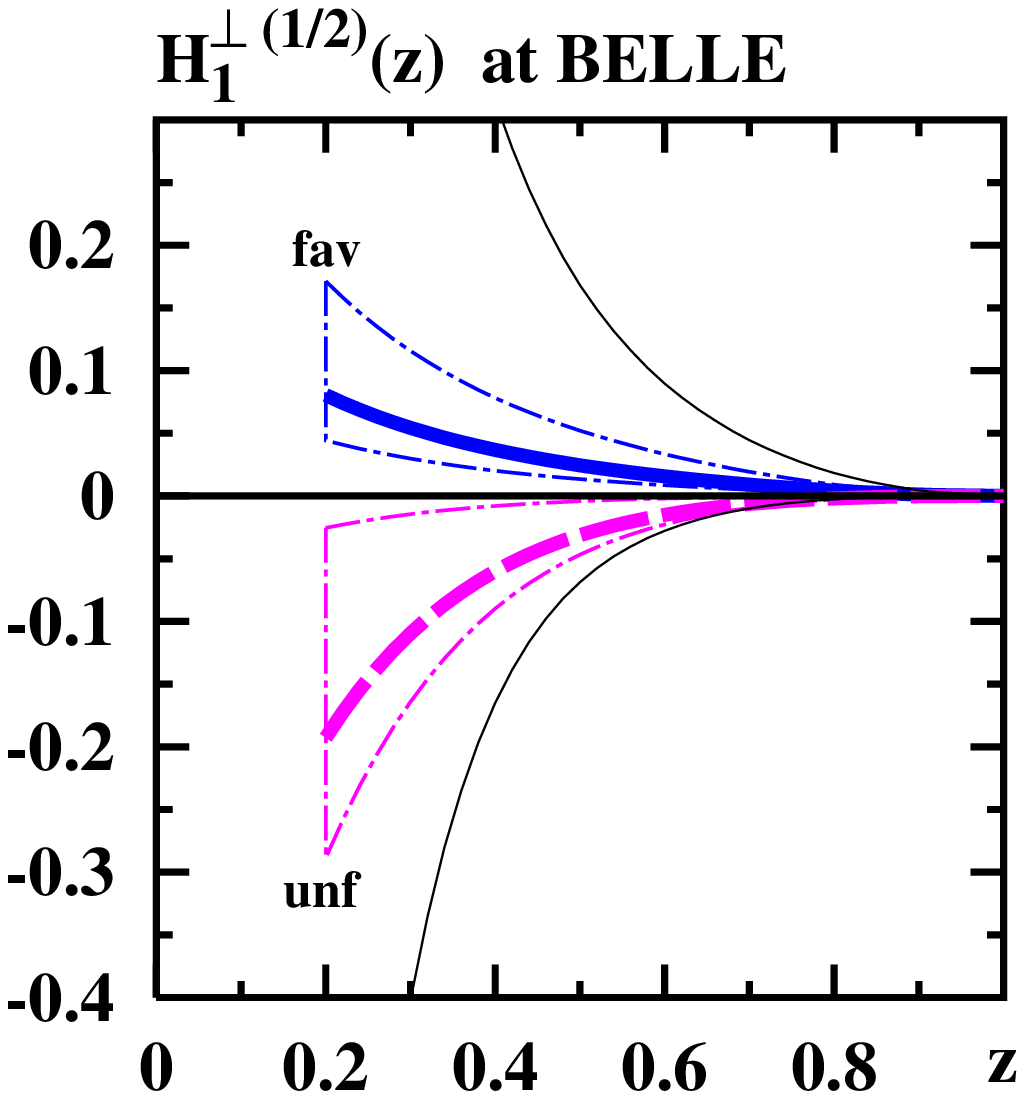}
\end{flushright}
\vspace{-5mm}
\begin{flushright}
\begin{minipage}{1.98in}
{\footnotesize {\bf FIGURE 2.}
Collins FF $H_1^\aH(z)$ needed to explain BELLE data \cite{Abe:2005zx}.}
\end{minipage}
\end{flushright}
\end{wrapfigure}
\addtocounter{figure}{1}       

\paragraph{3. Collins effect in $e^+e^-$}
The specific $\cos2\phi$ dependence of the cross section 
(\ref{Eq:A1-in-e+e}) could arise also from hard gluon 
radiation or detector acceptance effects. These effects, 
being flavor independent, cancel out from the double ratio 
of $A_1^U$, where both hadrons $h_1h_2$ are pions of unlike sign, 
to $A_1^L$, where $h_1h_2$ are pions of like sign, i.e.
\be
\label{Eq:double-ratio}
\frac{A_1^U}{A_1^L}\approx 1 + \cos(2\phi_1)P_1(z_1,z_2)\;.
\ee

In order to describe the BELLE data \cite{Abe:2005zx} we 
have chosen the Ansatz and obtained the best fit
\be
\label{Eq:best-fit-BELLE}
H_1^\aH(z) = C_a \,z\,D_1^a(z),\;\;\;
C_{\rm fav}=0.15,\;\;\;C_{\rm unf}=-0.45,
\ee
shown in Fig.~2 with 1-$\sigma$ error band (the errors are correlated).
Other Ans\"atze gave less satisfactory fits.

Notice that azimuthal observables in $e^+e^-$-anni\-hilation are 
bilinear in $H_1^\aH$ and therefore symmetric with respect to the 
exchange of the signs of $H_1^{\rm fav}$ and $H_1^{\rm unf}$. 
Thus in our Ansatz $P_1(z_1,z_2)$ is symmetric with respect to 
the exchange ${\rm sign}(C_{\rm fav})\leftrightarrow{\rm 
sign}(C_{\rm unf})$. (And not with respect to $C_{\rm 
fav}\leftrightarrow C_{\rm unf}$ as incorrectly remarked in 
\cite{Efremov:2006qm}.)

The BELLE data \cite{Abe:2005zx} unambiguously indicate that 
$H_1^{\rm fav}$ and $H_1^{\rm unf}$ have opposite signs,
but they cannot tell us which is positive and which is negative.
The definite signs in (\ref{Eq:best-fit-BELLE}) and Fig.~2
are dictated by SIDIS data \cite{Diefenthaler:2005gx}
(and our model \cite{Schweitzer:2001sr} with $h_1^u(x)>0$, see Sect.2).

\begin{figure}[b!]
\begin{tabular}{cccc}
\vspace{-1cm} &&& \cr
  \includegraphics[width=1.35in]{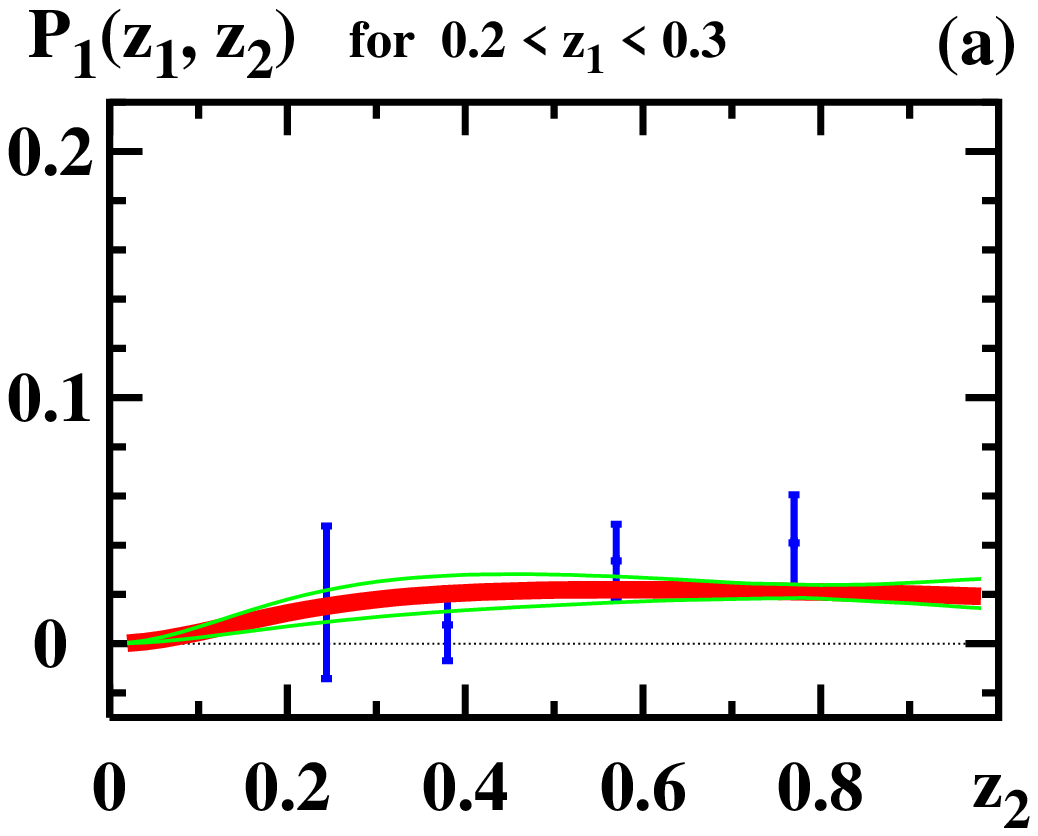}
& \includegraphics[width=1.35in]{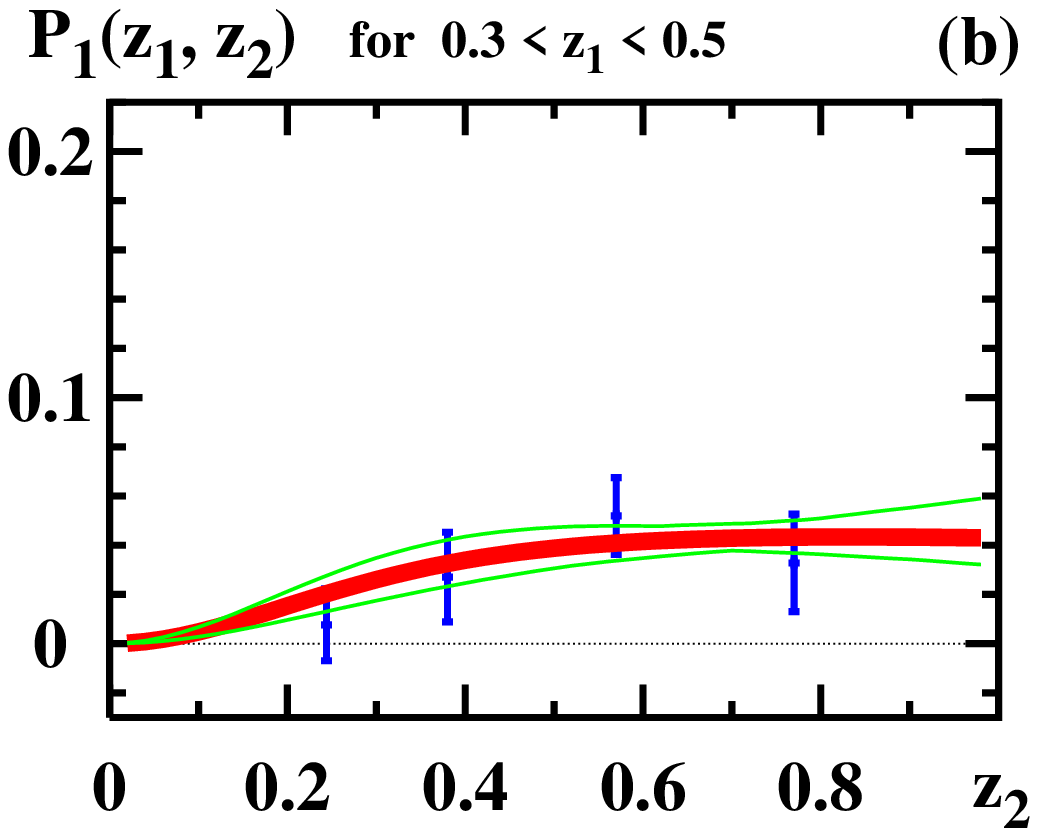}
& \includegraphics[width=1.35in]{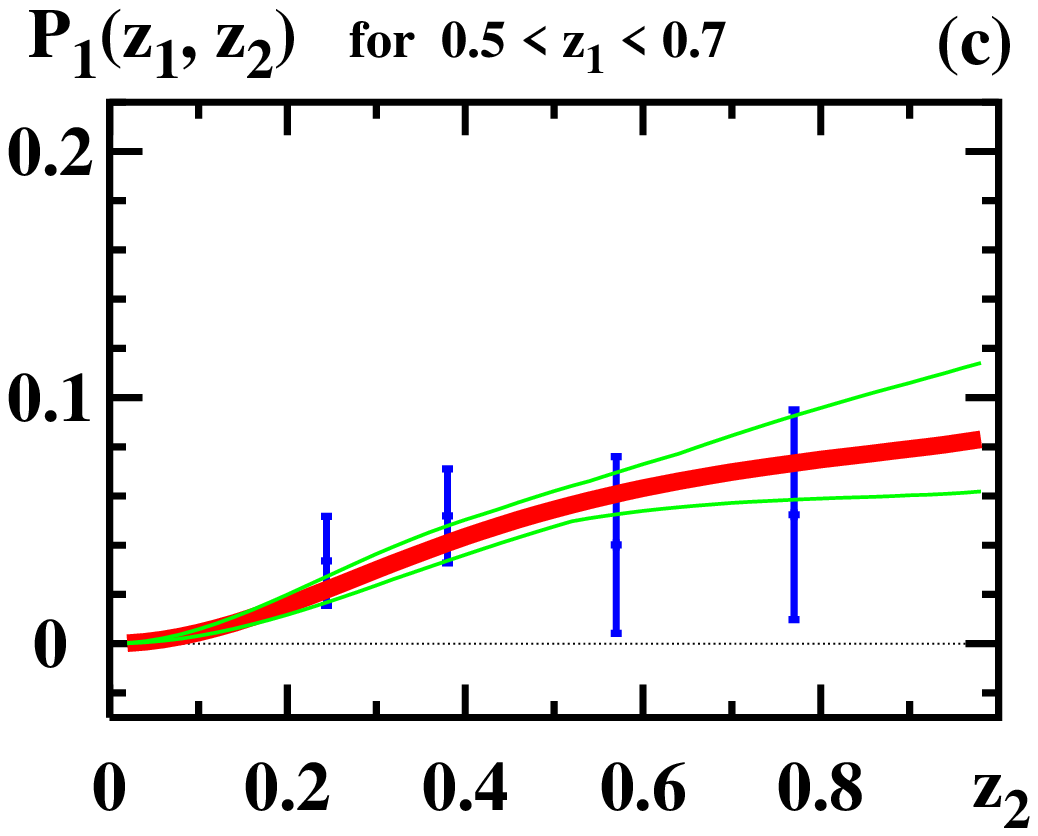}
& \includegraphics[width=1.35in]{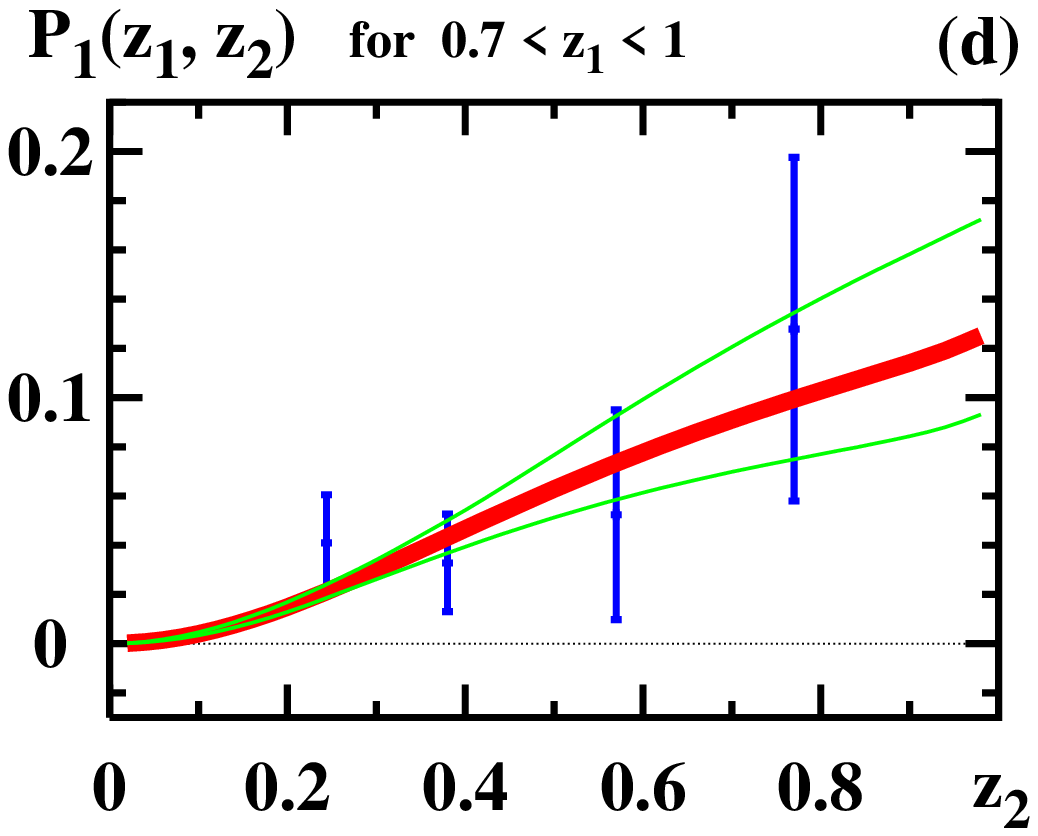} \cr
  \includegraphics[width=1.35in]{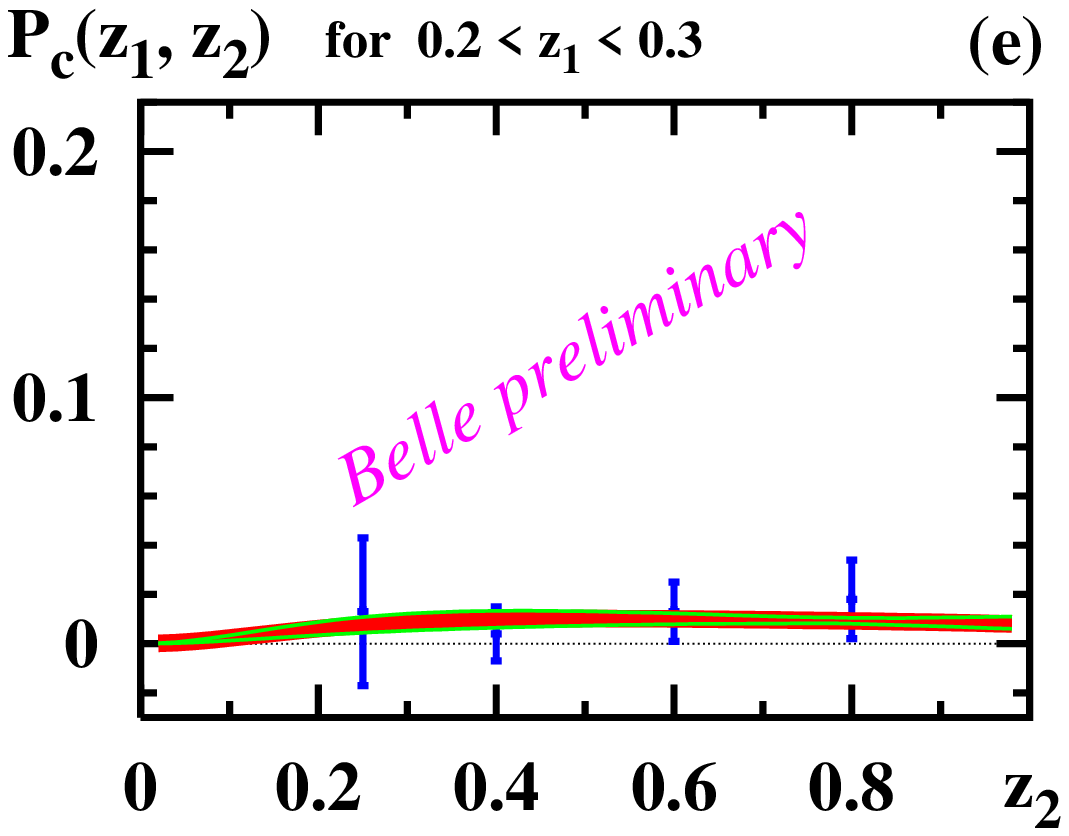}
& \includegraphics[width=1.35in]{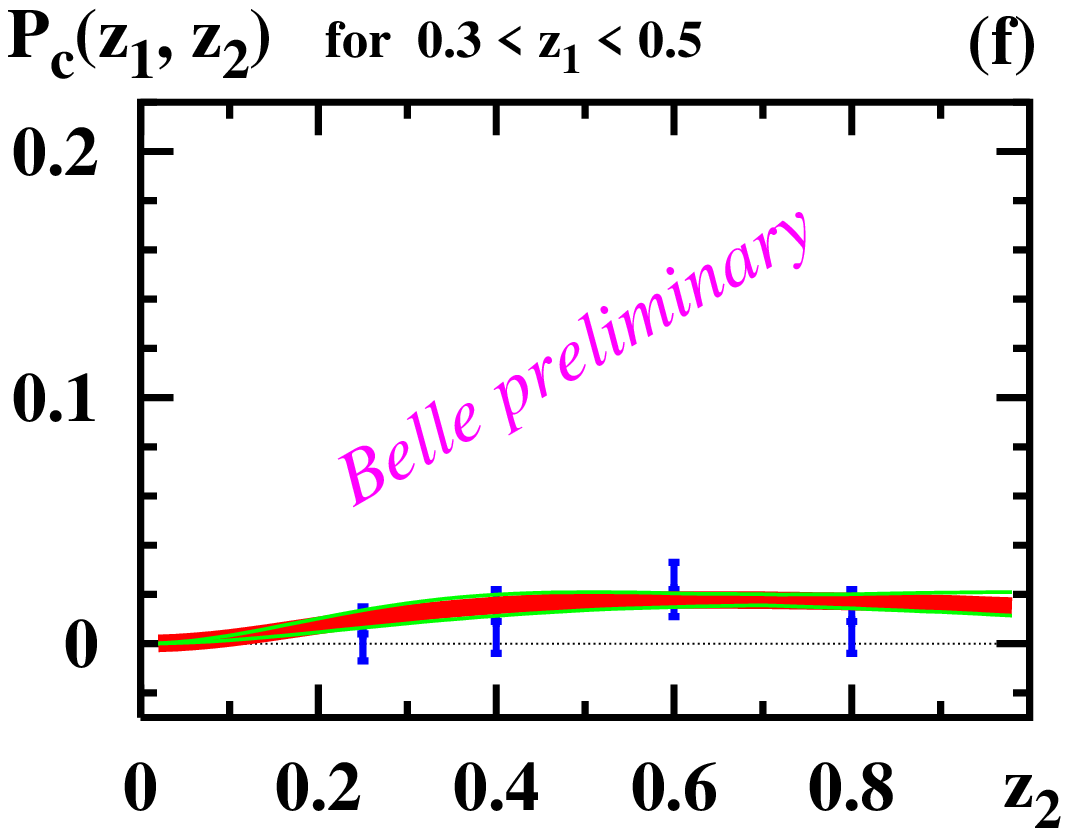}
& \includegraphics[width=1.35in]{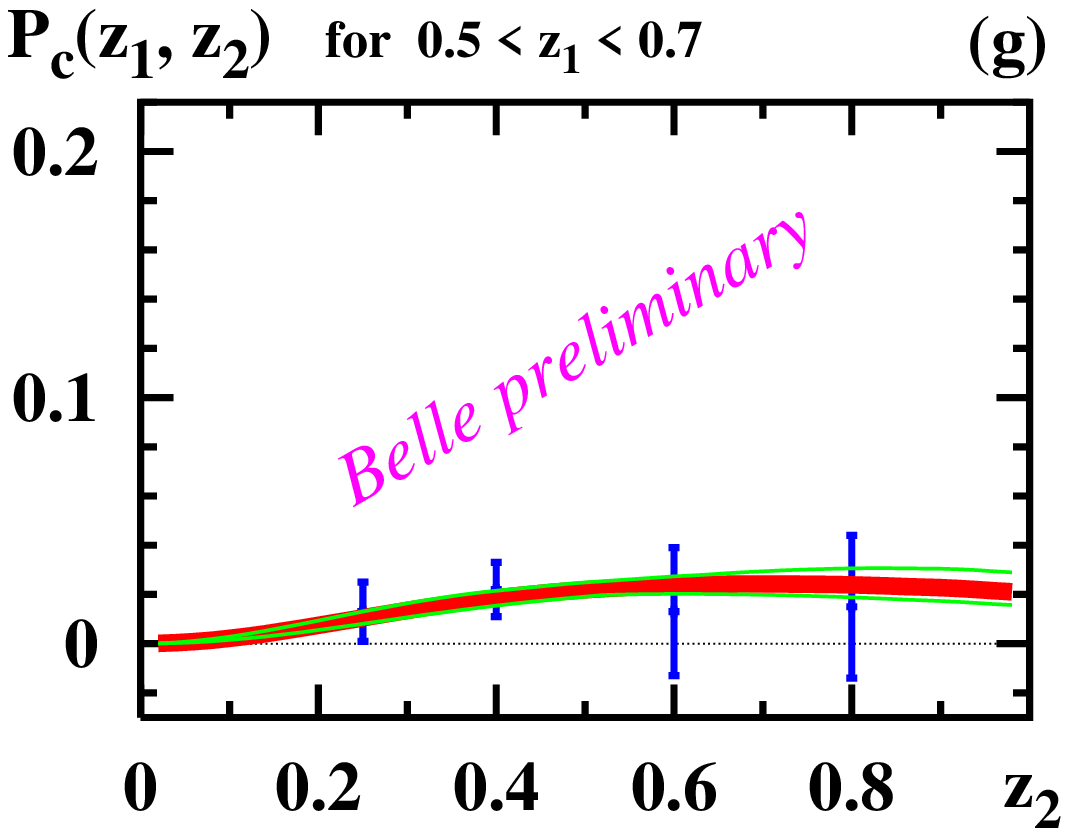}
& \includegraphics[width=1.35in]{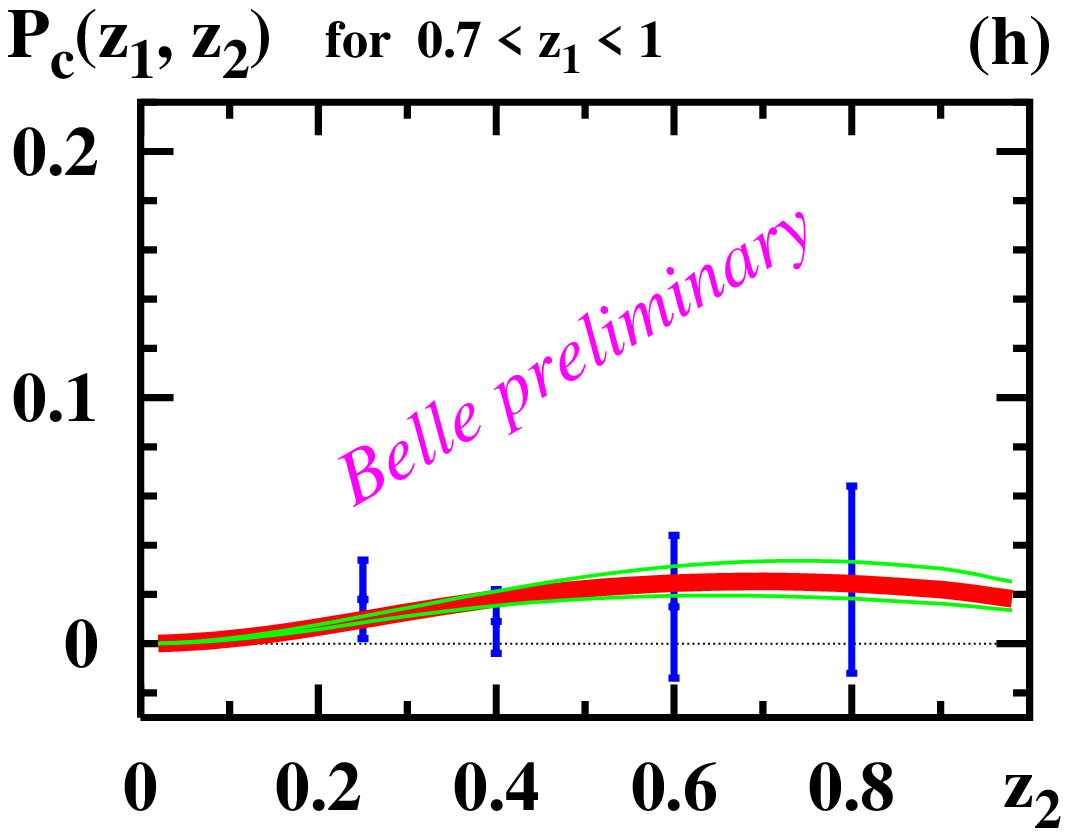}
\end{tabular}
\caption{\label{Fig6:BELLE}
    {\bf a-d}: 
    $\;\,P_1(z_1,z_2)$ as defined in
    Eq.~(\ref{Eq:double-ratio}) for fixed $z_1$-bins
    as function of $z_2$ vs.\ BELLE data \cite{Abe:2005zx}.
    {\bf e-h}: 
    The observable $P_C(z_1,z_2)$ defined analogously, see text,
    vs. preliminary BELLE data reported in \cite{Ogawa:2006bm}.}
\end{figure}

In Fig.~\ref{Fig6:BELLE}a-d the BELLE data \cite{Abe:2005zx} are 
compared to the theoretical result for $P_1(z_1,z_2)$ obtained on 
the basis of the best fit shown in 
Fig.~\ref{Fig5:BELLE-best-fit}2b. 

%
%

Most interesting recent news are the preliminary BELLE data 
\cite{Ogawa:2006bm} for the ratio of azimuthal asymmetries of 
unlike sign pion pairs, $A_1^U$, to all charged pion pairs, 
$A_1^C$. The new observable $P_C$ is defined analogously to $P_1$ 
in Eq.~(\ref{Eq:double-ratio}) as $A_1^U/A_1^C \approx
(1+\cos(2\phi)\,P_C)$. The fit (\ref{Eq:best-fit-BELLE}) ideally 
describes the new experimental points (see 
Figs.~\ref{Fig6:BELLE}e-h)!

\paragraph{4. BELLE vs.~HERMES}
In order to compare Collins effect in SIDIS at
HERMES \cite{Airapetian:2004tw,Diefenthaler:2005gx} and 
in $e^+e^-$-annihilation
at BELLE \cite{Abe:2005zx} we consider the ratios $H_1^a/D_1^a$ 
which might be less scale dependent. 
The BELLE fit in Fig.~\ref{Fig5:BELLE-best-fit}2 yields
in the HERMES $z$-range: 
\be\label{Eq:Apower-BELLE}
    \frac{\la 2H_1^{\rm fav}\ra}{\la D_1^{\rm fav}\ra}
    \biggr|_{\rm BELLE}\hspace{-9mm} = (5.3\cdots 20.4)\%,\quad
    \frac{\la 2H_1^{\rm unf}\ra}{\la D_1^{\rm
    unf}\ra} \biggr|_{\rm BELLE}\hspace{-9mm} =
    -(3.7\;\cdots\;41.4)\%  \;.
\ee
Comparing the above numbers (the errors are correlated!) to the
result in Eq.~(\ref{Eq:Apower-HERMES}) we see
that the effects at HERMES and at BELLE are compatible. The
central values of the BELLE analyzing powers seem to be
systematically larger but this could partly be attributed to
evolution effects and to the factor $B_{\rm G}<1$ in
Eq.~(\ref{Eq:Apower-HERMES}).
%
\begin{wrapfigure}[11]{R}{2.8in}
\vspace{-7mm}
\begin{flushright}
\includegraphics[width=1.25in]{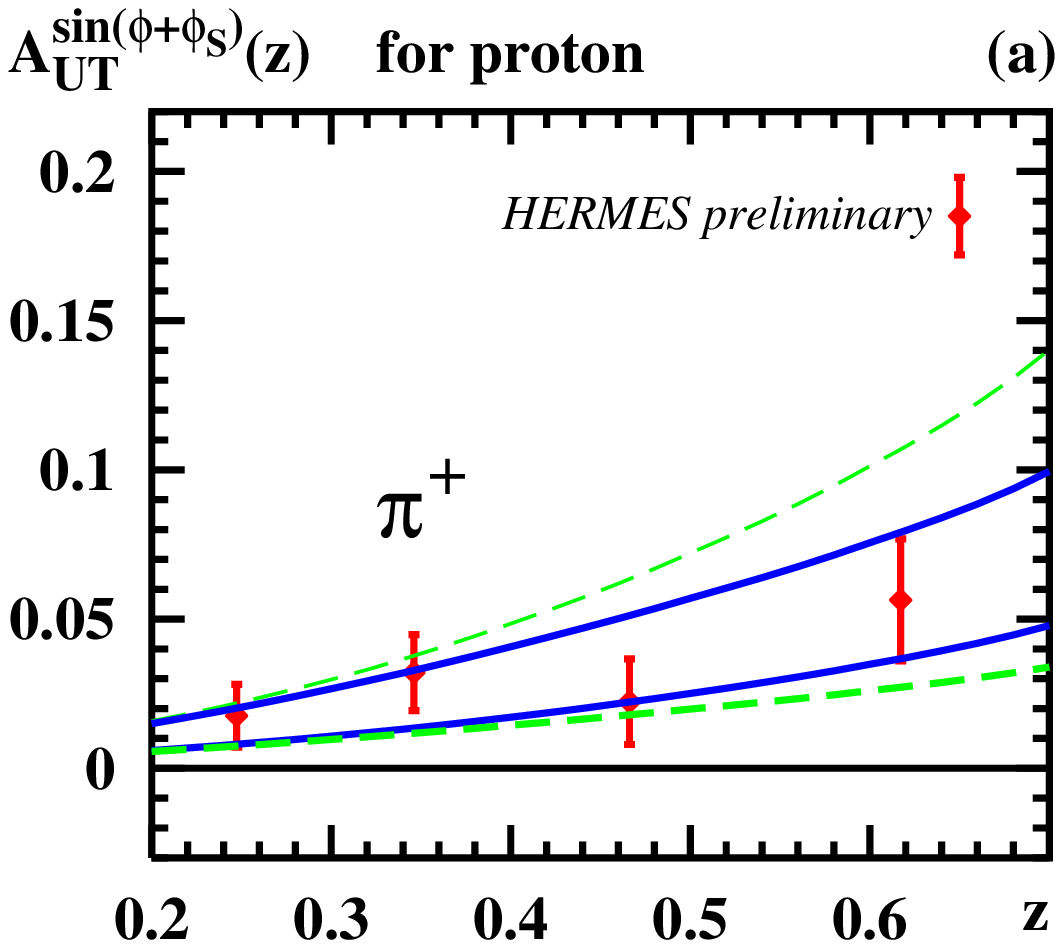}
\includegraphics[width=1.25in]{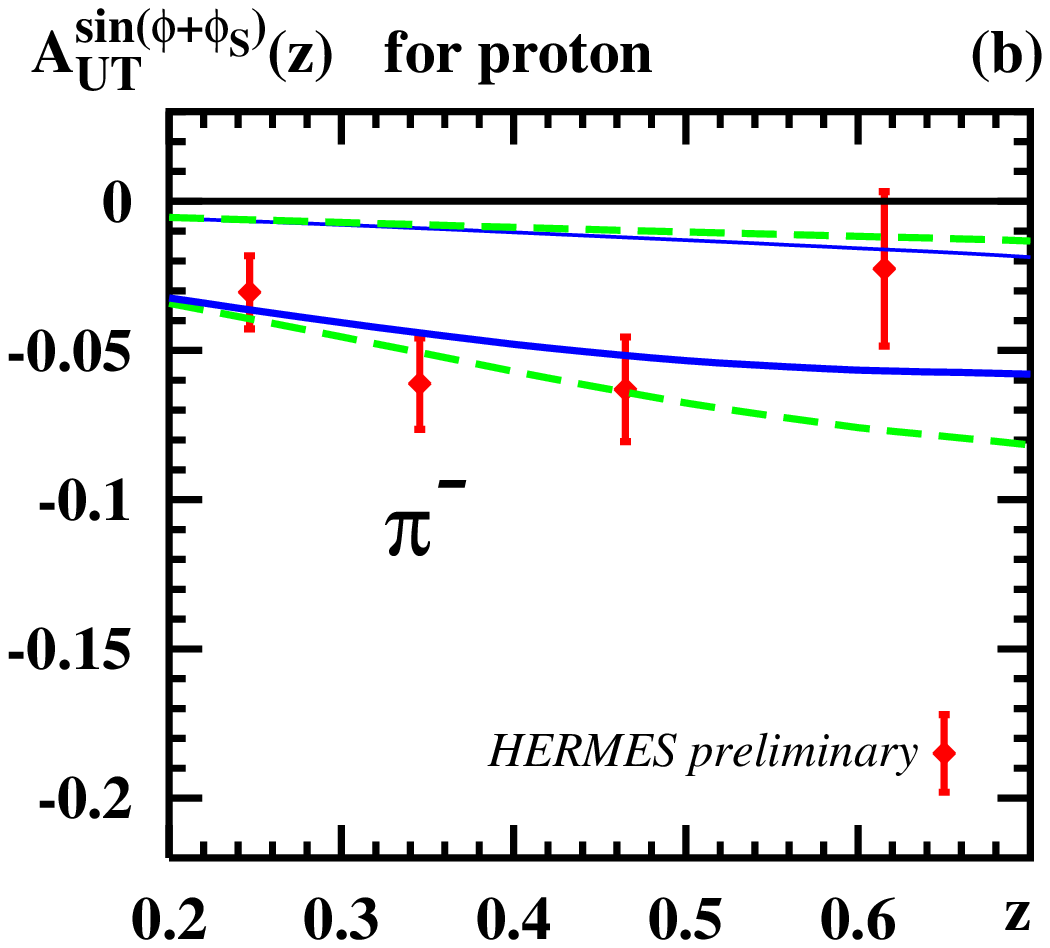}\\
\end{flushright}

\vspace{-5mm}
\begin{flushright}
\begin{minipage}{2.5in}
{\label{Fig7:HERMES-AUT-z-from-BELLE}
\footnotesize {\bf FIGURE~4.}
    The Collins SSA $A_{UT}^{\sin(\phi+\phi_S)}(z)$ as function
    of $z$. The theoretical curves are based on the fit of
    $H_1^a(z)$ to the BELLE data under the assumption
    (\ref{Eq:assume-weak-scale-dep}).
    }
\end{minipage}
\end{flushright}
\end{wrapfigure}
%

By assuming a weak scale-dependence also for the $z$-dependent ratios 
\be\label{Eq:assume-weak-scale-dep}
    \frac{H_1^\aH(z)}{D_1^a(z)}\biggr|_{\rm BELLE}
    \approx\;
    \frac{H_1^\aH(z)}{D_1^a(z)}\biggr|_{\rm HERMES}
\ee
and considering the 1-$\sigma$ uncertainty of the BELLE fit in 
Fig.~2 and the sensitivity to unknown Gaussian widths of 
$H_1^a(z)$ and $h_1^a(x)$, c.f.\  Footnote~1 and 
Ref.~\cite{Efremov:2006qm}, one obtains also a satisfactory 
description of the $z$-dependence of the SIDIS HERMES data 
\cite{Diefenthaler:2005gx}, see 
Fig.~\ref{Fig7:HERMES-AUT-z-from-BELLE}4.

These observations allow --- within the accuracy of the first data 
and the uncertainties of our study --- to draw the conclusion that 
it is, in fact, the same Collins effect at work in SIDIS 
\cite{Airapetian:2004tw,Diefenthaler:2005gx,Alexakhin:2005iw} 
and in $e^+e^-$-annihilation \cite{Abe:2005zx,Ogawa:2006bm}.
Estimates indicate that the early preliminary DELPHI result 
\cite{Efremov:1998vd} is compatible with
these findings, 
see \cite{Efremov:2006qm} for details.

%
\begin{wrapfigure}[11]{HR}{2.25in} 
\vspace{-7mm}
\begin{center}
\includegraphics[width=1.7in]{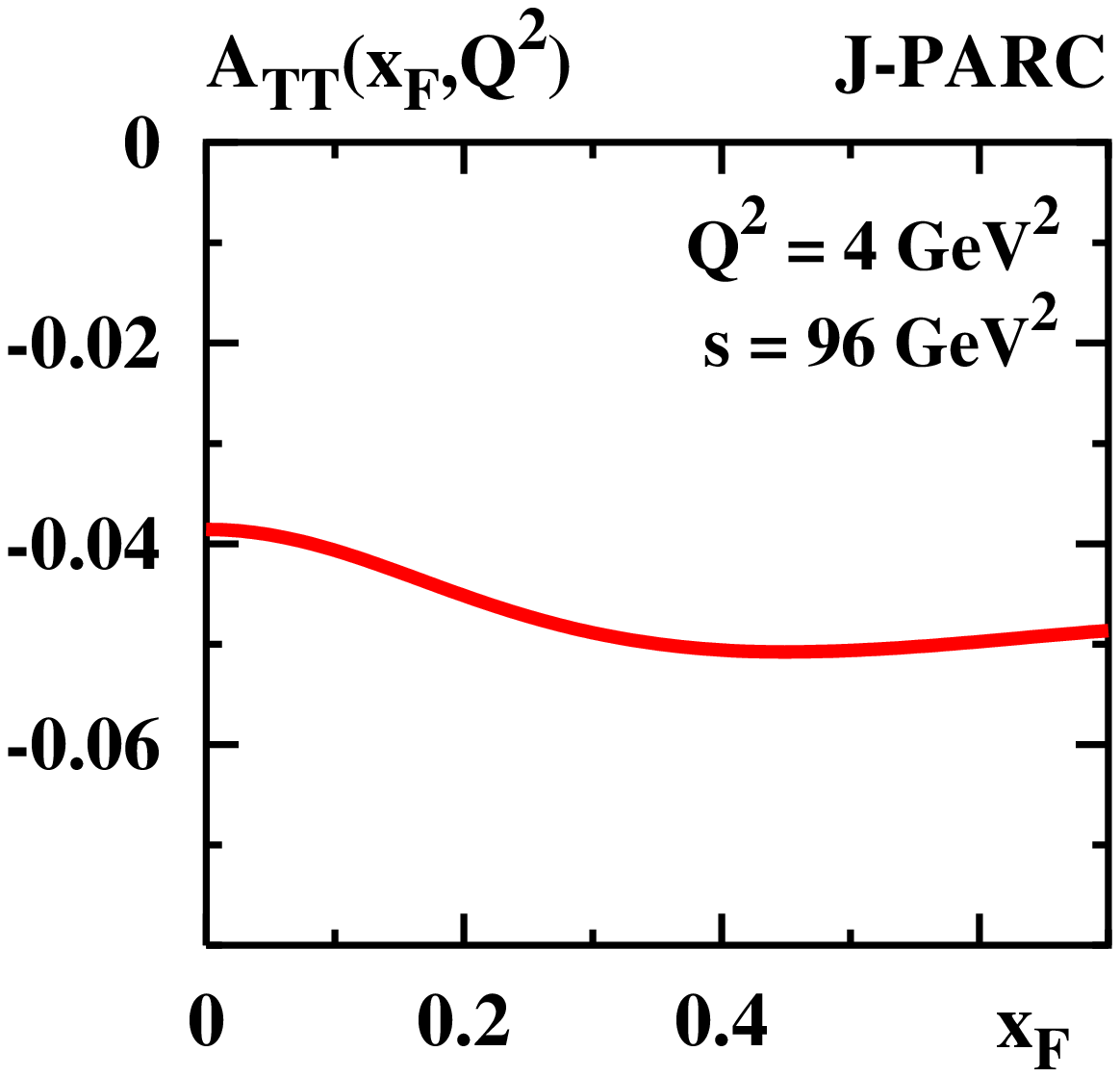}
\end{center}
\vspace{-5mm}
\begin{flushright}
\begin{minipage}{2in}
{\footnotesize {\bf FIGURE 5.}
Double spin asy\-mmetry $A_{TT}$ in DY, Eq.~(\ref{Eq:DY}),
vs.\ $x_F$ for the kinematics of J-PARC.
From \cite{in-progress}.}
\end{minipage}
\end{flushright}
\end{wrapfigure}
\addtocounter{figure}{1}       
\paragraph{5. Drell-Yan process}
The double-spin asymmetry observable in Drell-Yan (DY) 
lepton-pair production in proton-proton ($pp$) collisions 
is given in LO by
\be\label{Eq:DY}
    A_{TT}(x_F) = 
    \frac{\sum_a e_a^2 h_1^a(x_1) h_1^{\bar a}(x_2)}
         {\sum_a e_a^2 f_1^a(x_1) f_1^{\bar a}(x_2)}
\ee
where $x_F=x_1-x_2$ and $x_1x_2=\frac{Q^2}{s}$. In the kinematics 
of RHIC $A_{TT}$ is small and difficult to measure 
\cite{Bunce:2000uv}.

In the J-PARC experiment with $E_{\rm beam}=50\,{\rm GeV}$ 
$A_{TT}$ would reach $-5\,\%$ in the model 
\cite{Schweitzer:2001sr}, see Fig.~5, and could be measured 
\cite{J-PARC-proposal}. The situation is similarly promising in 
proposed U70-experiment \cite{Abramov:2005mk}.

Finally, in the PAX-experiment proposed at GSI \cite{PAX}
in polarized $\bar pp$ collisions one may expect 
$A_{TT}\sim(30\cdots50)\%$ \cite{Efremov:2004qs}.
There $A_{TT}\propto h_1^u(x_1) h_1^u(x_2)$ to a good approximation,
due to $u$-quark ($\bar u$-quark) dominance in the proton (anti-proton) 
\cite{Efremov:2004qs}.

\paragraph{6. Conclusions}
We studied the presently available data on the Collins effect.
Within the uncertainties of our study we find that the SIDIS data 
from HERMES \cite{Airapetian:2004tw,Diefenthaler:2005gx} and 
COMPASS \cite{Alexakhin:2005iw} on the Collins SSA from different
targets are in agreement with each other and with BELLE data on 
azimuthal correlations in $e^+e^-$-annihilations \cite{Abe:2005zx}.

The following picture emerges: favored and unfavored Collins FFs 
appear to be of comparable magnitude but have opposite signs, and 
$h_1^u(x)$ seems close to saturating the Soffer bound while the 
other $h_1^a(x)$ are presently unconstrained 
\cite{Efremov:2006qm}. 

These findings are in agreement with the most recent BELLE data 
\cite{Ogawa:2006bm} and with independent theoretical studies 
\cite{Vogelsang:2005cs}.

Further data from SIDIS (COMPASS, JLAB \cite{Chen:2005dq}, 
HERMES) and $e^+e^-$ colliders (BELLE) will help to refine and 
improve this first picture. 

The understanding of the novel functions $h_1^a(x)$ and 
$H_1^a(z)$ emerging from SIDIS and $e^+e^-$-annihilations, 
however, will be completed and critically reviewed only due to 
future data on double transverse spin asymmetries in the 
Drell-Yan process. Experiments are in progress or planned at 
RHIC, J-PARC, COMPASS, U70 and PAX at GSI.


\begin{theacknowledgments}
This work is  supported by BMBF (Verbundforschung), COSY-J\"ulich 
project, the Transregio Bonn-Bochum-Giessen, and is part of the 
by EIIIHT project under contract number RII3-CT-2004-506078. A.E. 
is also supported by RFBR grant 06-02-16215, by RF MSE 
RNP.2.2.2.2.6546 (MIREA) and by the Heisenberg-Landau Program of 
JINR.
\end{theacknowledgments}

\end{document}